\documentclass[%
 reprint,
superscriptaddress,
a4paper,
longbibliography,
amsmath,amssymb,
aps,prl,
numbers,
]{revtex4-1}

\usepackage[pdftex]{hyperref,graphicx}
\usepackage[usenames, svgnames]{xcolor}
\usepackage{amsfonts,amssymb,amsmath,mathrsfs}
\usepackage[separate-uncertainty=true]{siunitx}
\usepackage[english]{babel}
\usepackage[utf8]{inputenc}
\usepackage[T1]{fontenc}
\usepackage{xspace}

\newcommand{\figref}[2]{\hyperref[#1]{\getrefnumber{#1}(#2)}}

\hypersetup{
 pdftitle = {Spectral imaging of topological edge states in plasmonic waveguide arrays},
 pdfauthor = {Felix Bleckmann, Zlata Cherpakova, Stefan Linden, Andrea Alberti},
	colorlinks=true,linkcolor=blue,citecolor=blue,filecolor=blue,urlcolor=blue,pdfpagemode=UseNone,pdfstartview={XYZ null null 1.00}
}

\addto\captionsenglish{}
\addto\captionsenglish{}

\definecolor{DarkOrange}{rgb}{1,0.4,0.1}
\definecolor{LightOrange}{rgb}{0.7,0.45,0.}

\renewcommand\textemdash{\leavevmode\unskip\kern0.8pt\rule[0.19\baselineskip]{8pt}{0.4pt}\kern1pt\ignorespaces}

\begin{document}

\preprint{APS/123-QED}

\title{Spectral imaging of topological edge states in plasmonic waveguide arrays}

\author{Felix Bleckmann}
\affiliation{Physikalisches Institut, Rheinische Friedrich-Wilhelms-Universit\"at Bonn, Nu\ss{}allee 12, 53115 Bonn, Germany.}
\author{Zlata Cherpakova}
\affiliation{Physikalisches Institut, Rheinische Friedrich-Wilhelms-Universit\"at Bonn, Nu\ss{}allee 12, 53115 Bonn, Germany.}
\author{Stefan Linden}%
\affiliation{Physikalisches Institut, Rheinische Friedrich-Wilhelms-Universit\"at Bonn, Nu\ss{}allee 12, 53115 Bonn, Germany.}
\author{Andrea Alberti}%
\affiliation{Institut f\"ur Angewandte Physik, Rheinische Friedrich-Wilhelms-Universit\"at Bonn, Wegelerstraße 8, 53115 Bonn, Germany.}

\date{\today}

\begin{abstract}
We report on the observation of a topologically protected edge state at the interface between two topologically distinct domains of the Su-Schrieffer-Heeger model, which we implement in arrays of evanescently coupled dielectric-loaded surface plasmon polariton waveguides.
Direct evidence of the topological character of the edge state is obtained through several independent experiments:
Its spatial localization at the interface as well as the restriction to one sublattice is confirmed by real-space leakage radiation microscopy.
The corresponding momentum-resolved spectrum obtained by Fourier imaging reveals the midgap position of the edge state as predicted by theory.
\end{abstract}

\pacs{Valid PACS appear here}                \maketitle
\begin{figure}[t]
\centering
\includegraphics{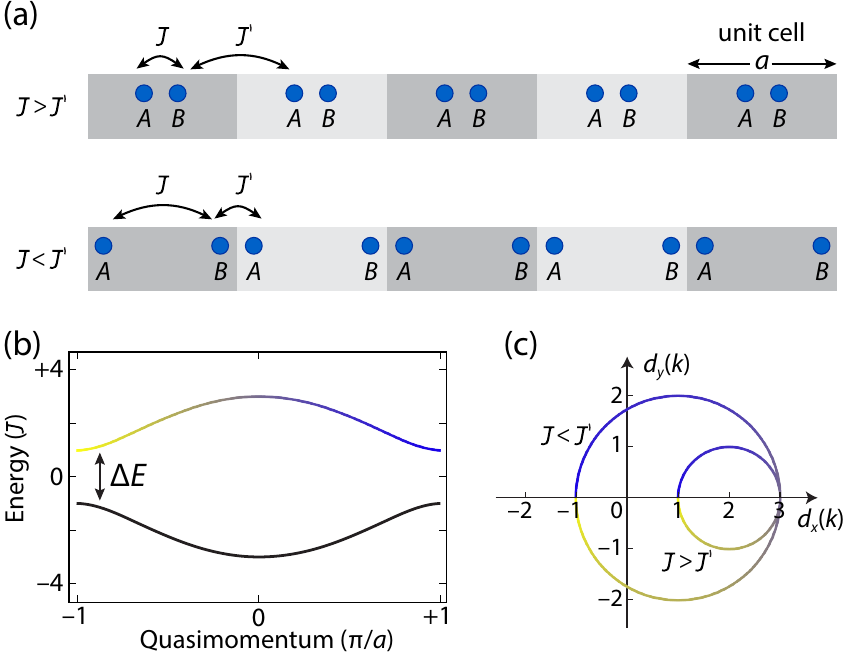}
\caption{(a) The two distinct dimerizations of the SSH model.
The intra and inter unit cell hopping amplitudes are characterized by $J$ and $J^\prime$, respectively.
(b) Band structure of the SSH model for $J^\prime = J/2$. (c) The $x$-$y$ components of the eigenstates $\mathbf{d}(k)$ for both dimerizations.
The same color scheme is used in (b) and (c) to indicate quasimomentum $k$ in the Brillouin zone.
}
\label{Fig1}
\end{figure}

Systems with non-trivial topological properties have attracted considerable interest since the discovery of the quantum Hall effect \cite{Klitzing:1980,Thouless:1982}. In particular topological insulators have been intensively studied in condensed matter physics~\cite{Hasan:2010,Qi:2011}. These materials behave as ordinary insulators in the bulk. However, at the boundary they exhibit topologically protected edge states, which in two dimensions (2D) can conduct unidirectional currents along the boundary without backscattering.
The prototypical system of a one-dimensional (1D) topological insulator is the Su-Schrieffer-Heeger (SSH) model~\cite{Su:1979}, i.e., a chain of identical sites coupled via alternating strong and weak bonds. 
The SSH model supports two different dimerizations characterized by distinct  topological invariants~\cite{ Delplace:2011,Atala:2013}, which depend on the choice of the unit cell (see Fig.~\ref{Fig1}).
As a consequence of the bulk-boundary correspondence principle~\cite{Asboth:2016}, a protected edge state is supported at each interface between the two different dimerizations.

Discrete photonic systems such as coupled waveguide arrays can show dynamics resembling quantum mechanical condensed matter phenomena. The basis for this is the mathematical equivalence between the time-dependent Schr\"odinger equation and the paraxial Helmholtz equation that describes the propagation of light \cite{Kogelnik:1996,Somekh:1973,Christodoulides:2003,Longhi:2009}.
Mapping the temporal dynamics of an electronic wave packet to the spatial evolution of the light field in a discrete photonic system thus allows us to experimentally study quantum mechanical evolutions.
Based on this approach, photonic topological insulators \cite{Lu:2014} consisting of helical dielectric waveguides \cite{Rechtsman:2013,Rechtsman:2013b}, 1D waveguide arrays \cite{Malkova:2009,Cheng:2015}, 1D quasicrystals \cite{Kraus:2012,Baboux:2016}, and discrete optical elements \cite{Kitagawa:2012} have been studied;
photonic systems not relying on the time-space mapping, such as coupled optical resonators \cite{Hafezi:2013}, and metamaterials \cite{Fleury:2016,Chen:2014}, have also been employed to study nontrivial topological systems.

While in 2D materials the topological protection of edge modes propagating along the interface separating topologically distinct domains is clearly assessed through the absence of backscattering,
in 1D systems, such as the SSH model and Majorana fermions in nanowires \cite{NadjPerge:2014}, evidence of the topological nature of edge states is much more elusive.
In 1D, the observation of a localized edge state is not sufficient to demonstrate its topological nature, since its appearance could in principle simply result from the broken translation symmetry (e.g., defect centers).

In this article, we report on the spectral imaging of the topologically protected edge state
of the SSH model implemented in a plasmonic waveguide array.
The employed detection technique uses leakage radiation microscopy \textemdash a distinct advantage of plasmonic systems \textemdash to experimentally identify the topological nature of the detected state through: (1) its spatial localization at the interface, (2) a direct probe of the underlying sublattice symmetry, and (3) its midgap position in the momentum-resolved spectrum \cite{Jackiw:1976, Su:1979,Asboth:2016}.
Related experiments observing a topological midgap edge mode in position space have been recently conducted with arrays of microwave resonators~\cite{Bellec:2014,Poli:2015}.

Before we address the experiments, we briefly summarize the properties of the SSH model. We start with the discussion of the bulk properties of an infinite chain of identical sites with alternating strong and weak bonds. The two topologically distinct dimerizations are depicted in Fig.~\figref{Fig1}{a}, where the strong (weak) bonds correspond to small (large) distances between two neighboring sites.
Considering only coupling among neighboring sites, the Hamiltonian of the SSH model for a given quasimomentum $k$ is:
\begin{equation}
	\nonumber
	\widehat{H}(k) = \left[ J+J^\prime \cos{(k \hspace{1pt}{a})} \right] \: \widehat{\sigma}_{x}+J^\prime \sin{(k \hspace{1pt}{a})} \: \widehat{\sigma}_{y} = \mathbf{d}(k) \cdot \widehat{\boldsymbol\sigma}, 
\end{equation}
where $a$ represents the lattice constant, $J$ and $J^{\prime}$ designate the hopping amplitudes inside and between unit cells. The components of the vector $\widehat{\boldsymbol\sigma} = \left[ \widehat{\sigma}_{x}, \widehat{\sigma}_{y}, \widehat{\sigma}_{z} \right]$ denote the Pauli matrices acting on the basis of the unit cell, i.e., the two sublattices formed by either sites A or sites B, see Fig.~\figref{Fig1}{a}.
The spectrum of $\widehat{H}(k)$ is composed of two energy bands  $E(k)=\pm \left| \mathbf{d}(k) \right|$, which are separated by a gap $\Delta E= 2|J-J^\prime|$, see Fig.~\figref{Fig1}{b}.
As $k$ is varied across the Brillouin zone $[-\pi/a,\pi/a]$ describing a closed loop in the parameter space, the vector $\mathbf{d}(k)$ representing the quantum state of the Bloch wavefunction forms a circle, see Fig.~\figref{Fig1}{c}.
Since $d_z(k)=0$, the circle is constrained to the x-y plane, meaning that the eigenstates in the bulk have equal weights on both sublattices.
However, the circle can either enclose or not the origin depending on the dimerizations (either $J>J'$ or $J<J'$).
Whether the origin is enclosed or not defines a topological invariant of the dimerization,
since one dimerization cannot be continuously transformed into the other without crossing the origin, i.e., closing the band gap.
Hence, the two dimerizations represent two different topological phases of the SSH model.
Moreover, the SSH Hamiltonian obeys chiral symmetry, meaning that a unitary operator $\widehat{\Gamma}$ exists fulfilling the condition $\widehat{\Gamma}\hspace{1pt} \widehat{H}\hspace{1pt}\widehat{\Gamma}^\dagger=-\widehat{H}$; from $d_z(k)=0$, it directly follows that \raisebox{0pt}[0pt][0pt]{$\widehat{\Gamma}$} is $\widehat{\sigma}_z$.
When topologically distinct domains are spatially connected, while preserving chiral symmetry, a bulk-boundary correspondence relates the difference of their topological invariants  to the number of topologically protected edge states at the interface~\cite{Hasan:2010}, namely one for the SSH model.
In addition, owing to chiral symmetry, this state has zero energy and it exhibits a vanishing amplitude on every second lattice site (sublattice symmetry).
Its zero-energy pinning is the underlying physical principle conferring topological protection on the edge state, since it prevents it from merging with the bulk continuum under a continuous deformation of the system's parameters.

\begin{figure*}[ht]
\centering
\includegraphics{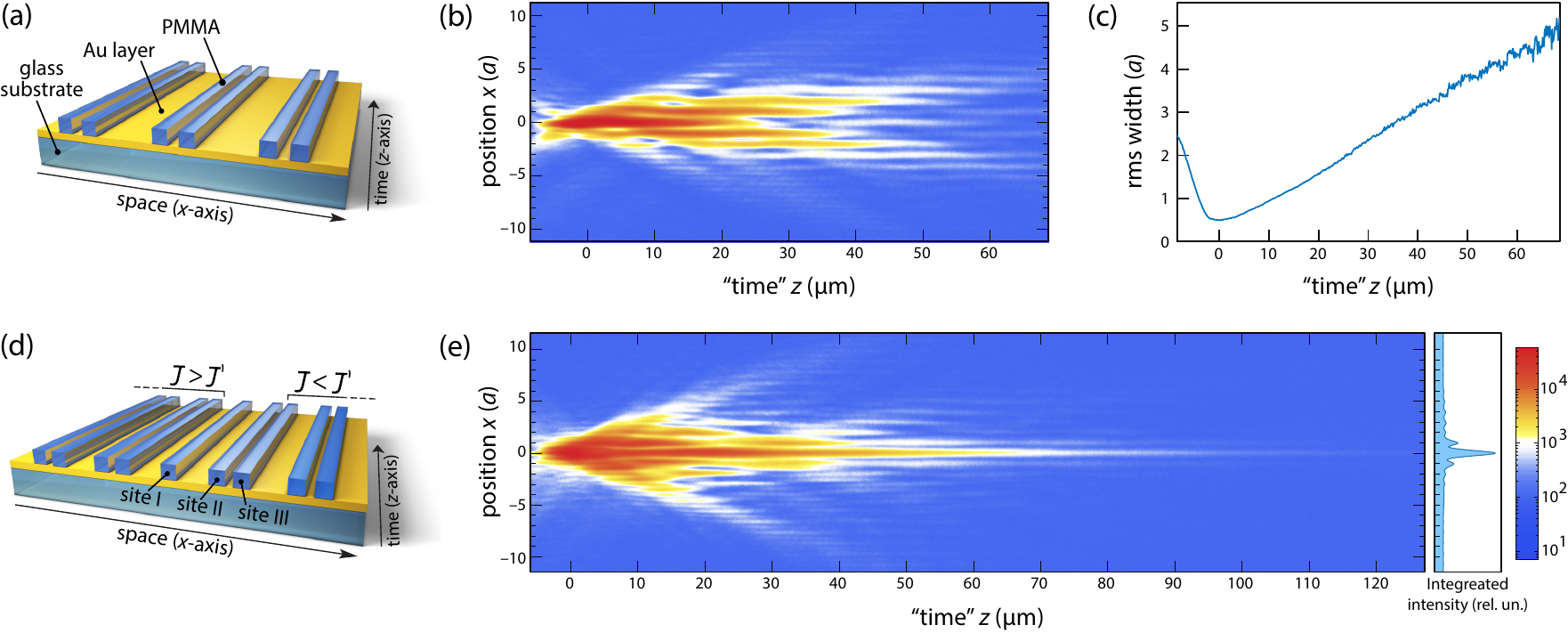}
\caption{(a) Waveguide array fabricated out of PMMA on top of a Cr- and Au-coated glass substrate.
Alternating center-to-center separations, \SI{600}{\nano \meter} and \SI{1000}{\nano \meter}, implement the bulk SSH model.
(b) Leakage radiation image of the spatial evolution of SPPs propagating along the structure depicted in (a).
The fading intensity along the $z$-axis is induced by radiation losses and absorption (propagation length $\approx \SI{16}{\micro\meter}$).
The color scale representing the SPP intensity is the same as in (e).
(c) rms width of the intensity distribution in (b), demonstrating ballistic expansion.
Ballistic expansion for $z<0$ correponds to free SPPs propagating at the gold-air interface.
(d) Plasmonic waveguide array incorporating a topological defect where the long separation is repeated twice.
Three different excitation sites, I, II, and III, are highlighted.
(e) Leakage radiation image of SPPs propagating along the structure depicted in (d) with excitation at site~I. A pronounced central feature is visible around the topological defect at $x=0$. The ballistically expanding background results from SSH bulk modes that are also excited.
The inset shows the intensity distribution integrated along the $z$-axis in the region $\SI{50}{\micro \meter}<z<\SI{130}{\micro \meter}$.
}
\label{Fig2}
\end{figure*}

We employ arrays of evanescently coupled dielectric-loaded surface plasmon polariton (SPP) waveguides \cite{Holmgaard:2007,Grandidier:2008} to realize the SSH model.
As for coupled dielectric waveguides \cite{Somekh:1973,Christodoulides:2003,Longhi:2009}, we rely on the foregoing equivalence between the Schrödinger equation and the Helmholtz equation to describe the propagation of SPPs in waveguide arrays.
The arrays are fabricated by negative-tone grey-scale electron beam lithography \cite{Block:2014} on top of a chromium (\SI{10}{\nano \meter}) and gold (\SI{60}{\nano \meter}) coated glass substrate.
The  waveguides consist of polymethylmethacrylate (PMMA) ridges with a width of \SI{250}{\nano \meter} and a height of \SI{140}{\nano \meter}.
We realize strong and weak bonds as in the SSH model by alternating different separations, \SI{600}{\nano \meter} and \SI{1000}{\nano \meter}, between neighboring  waveguides, as depicted in Fig.~\figref{Fig2}{a}, resulting in $a=\SI{1600}{\nano\meter}$.
These geometrical parameters  ensure single mode operation of the waveguides and sufficient coupling among them for the vacuum wavelength $\lambda=\SI{980}{\nano \meter}$.
In all experiments, this excitation wavelength is chosen as a good trade-off between the absorption losses in gold and the decreasing camera efficiency for longer wavelengths.

SPPs are excited by focusing a transverse-magnetic polarized laser beam with wavelength $\lambda$ onto a PMMA grating coupler, which is situated on top of the first $\SI{5}{\micro\meter}$ of selected  waveguides (not shown in the figure).
The SPP evolution in the  waveguides is monitored by leakage radiation microscopy~\cite{Hecht:1996,Drezet:2008}.
An oil immersion objective lens with a numerical aperture (NA) of 1.49 collects the fraction of the excited SPPs that leaks through the thin gold film and coherently couples to freely propagating modes in the glass substrate.
The large NA allows us to resolve spatial structures as small as $\SI{500}{\nano\meter}$ [the modulation transfer function equals 0.2 at the spatial frequency of $1/(\SI{500}{\nano\meter})$].
After blocking the directly transmitted laser beam via angular filtering~\cite{Drezet:2006}, the remaining radiation is detected with a CMOS camera (Andor Zyla).
Momentum-resolved spectra of the SSH model are recorded by imaging the back-focal plane of the microscope objective lens onto the camera.

To provide a reference measurement, we study the spatial evolution of an initially localized plasmonic wave packet in the bulk of the SSH model, see Fig.~\figref{Fig2}{b}.
We excite SPPs in a single  waveguide through the grating coupler.
As the SPPs propagate along the waveguides (\mbox{$z$-axis}), the plasmonic field is coherently transferred to the neighboring waveguides.
This results in a characteristic interference pattern, which exhibits a linear increase of its width with the propagation distance (ballistic spreading), as visible in Fig.~\figref{Fig2}{c}.
 This transport behavior is analogous to the temporal evolution of electronic wave packets tunneling along a chain of lattice sites with alternating coupling strengths.

As we are interested in the physics at the boundary between topologically distinct domains, we excite SPPs in a  waveguide array fabricated with an interface between the two distinct dimerizations of the SSH model, as depicted in Fig.~\figref{Fig2}{d}.
Figure \figref{Fig2}{e} shows the measured spatial evolution in the case in which the excitation laser is coupled to the central waveguide (excitation site I).
In stark contrast to the reference measurement discussed above, a large fraction of the intensity stays confined to the interface region.
This aspect appears particularly evident if we consider the intensity distribution integrated along the propagation direction [inset of Fig.~\figref{Fig2}{e}], which displays a highly pronounced peak in the center.
A close inspection reveals further side peaks at the neighboring unit cells, whose intensities rapidly decay with the distance from the center.
We interpret these experimental observations as a clear evidence of an edge state localized at the interface, whose spatial mode has a large overlap with the excitation field.

\begin{figure}[b]
\centering
\includegraphics{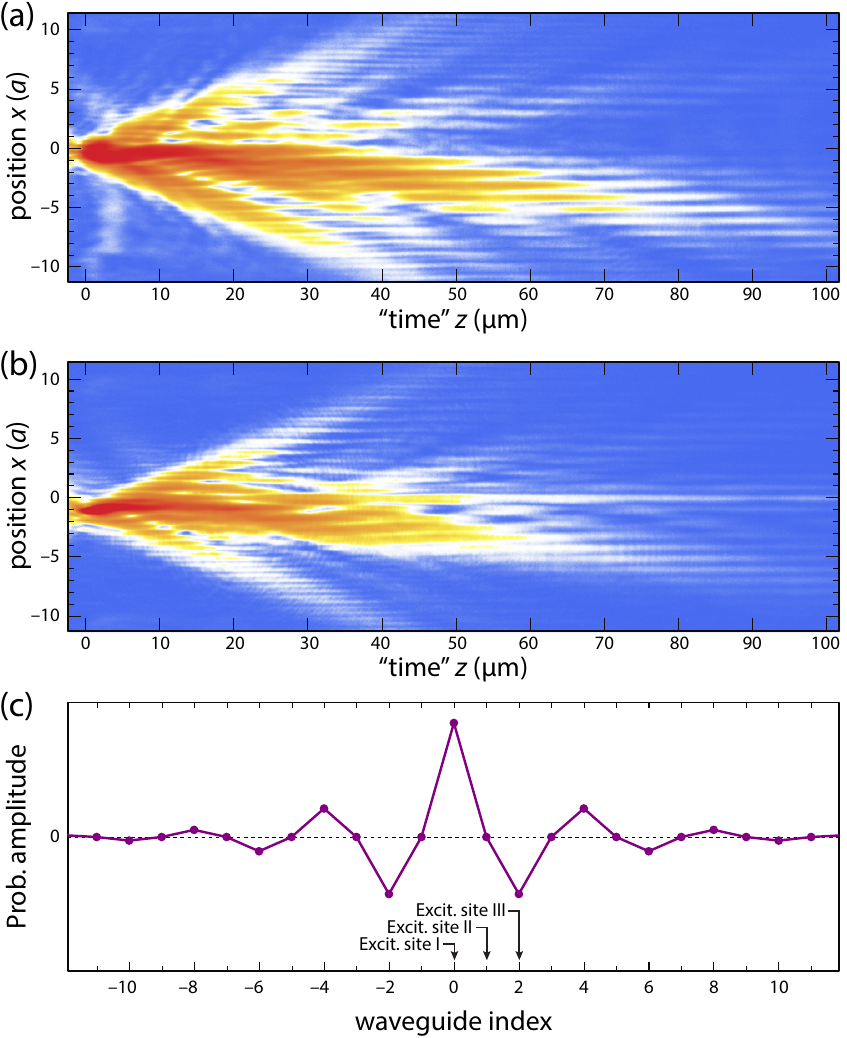}
\caption{
Leakage radiation image of propagating SPPs, with SPPs excited at the excitation site II (a) and excitation site III (b). The color scale is the same as in Fig.~\ref{Fig1}.
(c) Probability amplitude of the topologically protected edge state as predicted by the SSH theory for the waveguide geometry in Fig.~\figref{Fig2}{d}.
}
\label{Fig3}
\end{figure}

However, observing a localized state does not suffice to conclude that it is indeed the topologically protected edge state predicted by the SSH model.
To fill this void, we look for other experimental signatures pointing to the topological nature of the edge state, by conducting two complementary experiments.
A first hint is found by closely examining the intensity distribution shown in the inset of Fig.~\figref{Fig2}{e}, which, despite the finite optical resolution of our imaging system, reveals that the edge state is concentrated only on one sublattice.
To provide even stronger evidence, we repeat the experiment presented in Figs.~\figref{Fig2}{d-e}, however, with the excitation site shifted either by one waveguide (excitation site II) or by two waveguides (excitation site III).
For excitation site II, the recorded intensity distribution shown in Fig.~\figref{Fig3}{a} only displays ballistic spreading resulting from excited bulk modes. The edge state is not excited in this case. In contrast, the localized state can be clearly observed again, in addition to the bulk modes, if the waveguide array is excited at site III, see Fig.~\figref{Fig3}{b}.
These findings allow us to validate the SSH model's prediction that the topologically protected edge state has vanishing amplitudes on every other site, as displayed in
Fig.~\figref{Fig3}{c} \cite{Schomerus:2013}.

\begin{figure}[t]
\centering
\includegraphics{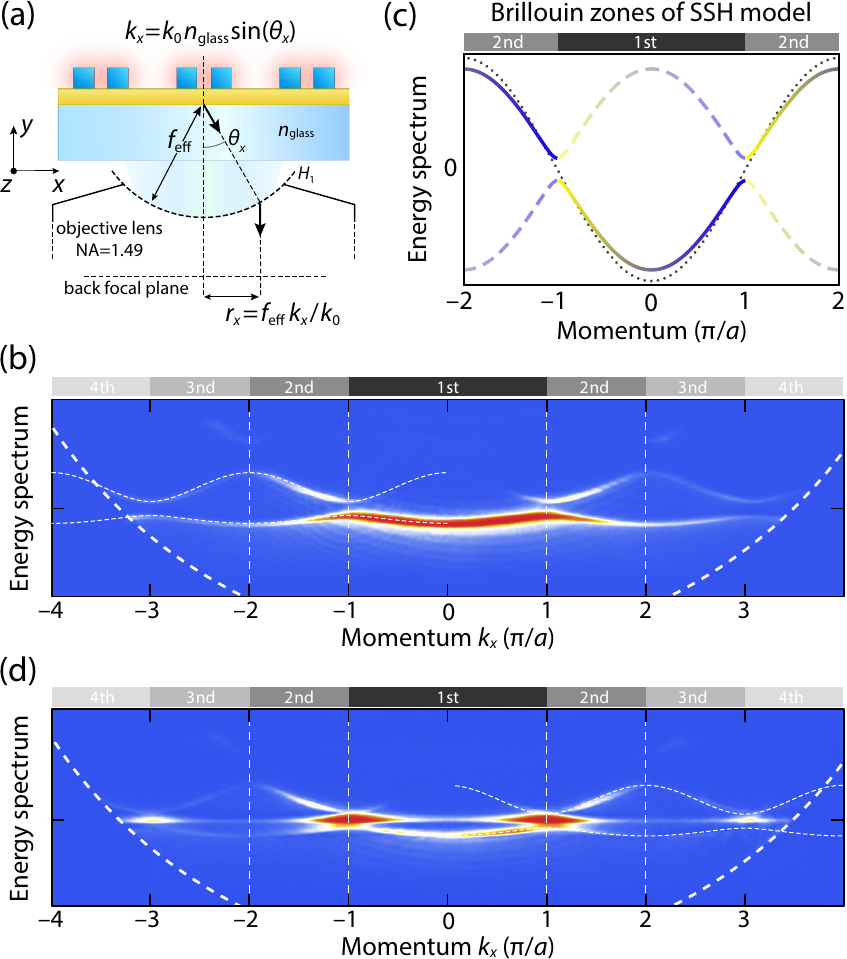}
\caption{(a) Measurement scheme for recording momentum-resolved spectra of SPPs via Fourier imaging of leakage radiation.
Leakage radiation preserves the SPP momentum component parallel to the glass substrate of index $n_\text{glass}$.
Principal plane $\text{H}_1$ and  sine condition are also shown.
(b)~Momentum-resolved spectrum of SPPs propagating in the bulk, see Fig.~\figref{Fig2}{a}, acquired by Fourier imaging of the leakage radiation (similar color scale as in Fig.~\ref{Fig1}).
As explained in the text, we interpret the momentum distribution along the $y$-axis as the energy spectrum of the SSH model; the vertical units can be gauged taking into account that the dashed circle denotes the maximum detected momentum ($k_0\hspace{1.5pt}\text{NA}$).
Thin dashed lines are the expected energy bands for the SSH model extended to include next-nearest neighbor coupling $J''$ ($J'\approx 0.5\,J$ and $J''\approx 0.1\,J$).
(c)~The dotted curve is the energy band of a simple lattice ($J=J'$, lattice constant $a/2$), the solid and dashed curves are the energy band in the presence of a weak imbalance ($J\neq J'$), see text. We use the same color scheme as in Fig.~\ref{Fig1}.
(d)~Like~(b) but for SPPs excited at the topological defect (site~I), see Fig.~\figref{Fig2}{d}.
}
\label{Fig4}
\end{figure}

The second experiment tests a complementary property of the edge state of the SSH model by directly measuring its spectral properties.
This is achieved by imaging the back-focal plane of the  objective lens (Fourier imaging).
For an aplanatic objective lens as ours, the so-called sine condition relates  the spatial intensity distribution in the back-focal plane of the objective to the angular distribution of the collected light.
Through this relation, the SPP wave vector $\mathbf{k}_\mathrm{SPP}=(k_x,k_z)$ is directly mapped to a given position $\mathbf{r}$ in the back-focal plane, $\mathbf{r} = f_\text{eff}\hspace{1pt} \mathbf{k}_\mathrm{SPP}/k_0$, where $k_0$ is the vacuum wavenumber and $f_\text{eff}$ is the effective focal length, as shown in Fig.~\figref{Fig4}{a}.
Hence, imaging the back-focal plane gives us direct access to the $k_x$ and $k_z$ momentum distribution of SPPs in the waveguide arrays.
Based on the time-space mapping between the SSH model and  waveguides, we interpret the $k_x$ distribution (transverse direction) as the momentum distribution and the $k_z$ distribution (SPP propagation direction) as the energy spectrum.

Figure~\figref{Fig4}{b} provides the reference measurement of the momentum-resolved bulk spectrum of the SSH model for the geometry depicted in Fig.~\figref{Fig2}{a}.
We excite a single waveguide in the bulk as for Fig.~\figref{Fig2}{b}. The recorded momentum-resolved bulk spectrum exhibits two cosine-like bands which are separated by a gap in the $k_z$-direction.
Based on the foregoing time-space mapping, we interpret these two bands as two energy bands of the SSH band structure shown in Fig.~\figref{Fig4}{c}.
However, instead of reconstructing the spectrum in the first Brillouin zone, our experimental technique gives us access to the full decomposition in momentum components in the higher Brillouin zones \footnote{Our technique is analogous to a time-of-flight detection of cold atoms after sudden release from an optical lattice.}.
The maximum detected momentum ($k_0\hspace{1.5pt}\text{NA}$) is determined by the NA of our objective lens (see dashed circle in the figure), which allows us to precisely gauge the scale of the momentum ($k_x$) and energy ($k_z$) axes.
Hence, with the knowledge of the lattice constant $a$, we can precisely identify the boundaries of the different Brillouin zones (vertical dashed lines).
Since we excite a single waveguide, all quasimomenta are occupied, as shown by the broad energy band distribution along the $k_x$-axis.
To understand the observed intensity distribution within the bands in more detail, it is instructive to first consider the case of a bulk 1D system with identical separations $a/2$ between waveguides, and hence same coupling, $J=J'$.
This results in a single cosine-like band, as indicated by the dotted line in Fig.~\figref{Fig4}{c}, with a first Brillouin zone twice as large ($4\pi/a$).
For this simple lattice, one expects a uniform occupation of this single band.
If every second site is shifted by a small amount, thereby introducing an imbalance between $J$ and $J'$, we retrieve the Brillouin zones of the SSH model.
As a result of the small perturbation, the momentum-resolved energy spectrum acquires a gap in the $k_z$-direction at the boundaries between Brillouin zones, while the intensity distribution remains largely unchanged, that is, strong (weak) occupation of the band indicated by the solid (dashed) line in Fig.~\figref{Fig4}{c}.
This simple model explains qualitatively the observed intensity distribution in Fig.~\figref{Fig4}{b}.
We attribute the slightly different spectral widths of the two observed bands to a weak next-nearest neighbor coupling, which constitutes a small perturbation to the SSH model breaking its chiral symmetry.

Figure \figref{Fig4}{d} displays the momentum-resolved spectrum of the SSH model in the case in which we excite the  waveguide array in the proximity of the defect (excitation site I).
We observe an additional mode with midgap position, which agrees with the prediction of the SSH model of an edge state with topological protection.
This gives us a third strong evidence that the excited edge state is in fact of topological nature.
We explain the small deviation from the exact band-gap center in terms of a nonvanishing next-nearest neighbor coupling.
Moreover, we infer from the high intensity of the edge state at the zone boundary that the amplitudes of the edge state alternate on every second waveguide, with vanishing population on the other sublattice.
This gives a further confirmation of the observation in Fig.~\ref{Fig3} that the edge mode is restricted to one sublattice, and therefore that this is the topologically protected edge state of the SSH model.

\begin{figure}[t!]
\centering
\includegraphics{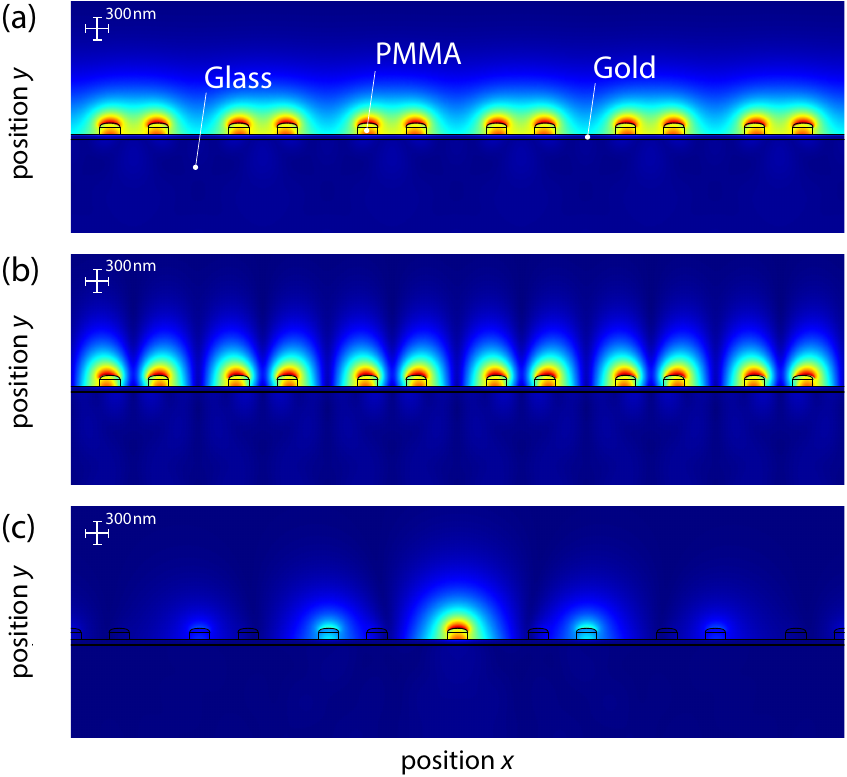}
\caption{Calculated SPP electric field distributions for solutions of the SSH model with quasimomentum $k{=}0$ for (a) the lower and (b) the upper band, with effective refractive indices $n_\text{eff}{=}1.085 + 0.004i$ and $n_\text{eff}{=}0.936 + 0.005i$, respectively. (c) Electric field distribution of the topological edge mode with $n_\text{eff}=1.044 + 0.004i$. The solid black lines depict the geometry of the waveguide arrays used in the experiments without (a,b) and with (c) the topological defect. The color scale shows the norm in arbitrary units.\vspace{-4mm}}
\label{Fig5}
\end{figure}

\begin{figure}[t!]
\centering
\includegraphics{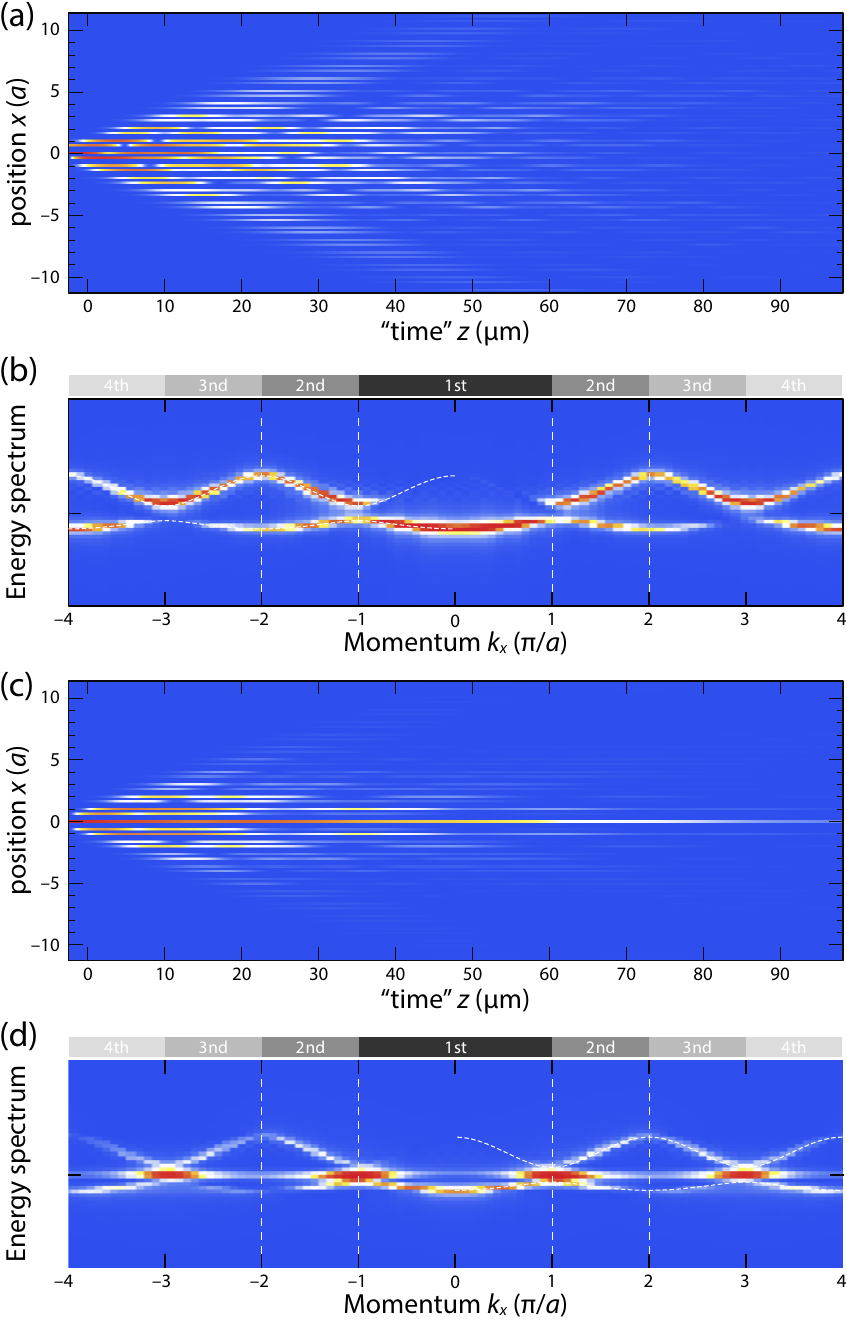}
\caption{Calculated SPP real space intensity distributions and momentum-resolved spectra. From top to bottom, the calculations correspond to the experiments presented in Figs.~\figref{Fig2}{b}, \figref{Fig4}{b}, \figref{Fig2}{e}, and \figref{Fig4}{d}, respectively. Units and color scales are the same as in the corresponding experimental graphs. The calculations are based on a finite element method in frequency space (see text).\vspace{-4mm}}
\label{Fig6}
\end{figure}

Alongside the interpretation of our experimental results based on the SSH model, we have also performed numerical calculations based on a finite element method in frequency space.
We solve the Helmholtz equation in the transverse plane (assuming translational symmetry along the waveguides) for plasmonic waveguide arrays with and without the topological defect in the center.
We thereby obtain the eigenmodes and the corresponding complex effective refractive indices, $n_\text{eff}=\beta\,\lambda/(2\pi)$, with $\beta$ being the propagation constant related to $k_z$, $\text{Re}(\beta){=}k_z$.
For the calculations, the array size is limited to 50 waveguides,
and the cross sections of the plasmonic waveguides as well as their separations are chosen according to the experiments.
Figure \figref{Fig5}{a} and \figref{Fig5}{b} show the calculated electric field distribution of bulk modes (quasimomentum $k{=}0$), demonstrating the bonding and antibonding orbitals for the lower and upper band, respectively.
The calculated electric field distribution of the topological edge mode is depicted in Fig.~\figref{Fig5}{c}, showing that the field has minima on every second waveguide.
Bulk modes, as well as the topological edge mode, have similar attenuation coefficients $\text{Im}(\beta)$, with small variations in the range of $\SI{30}{\percent}$.

The real space evolution of the SPP amplitudes in a given array is determined by decomposing the excitation field in terms of the eigenmodes, and by letting each eigenmode acquire a phase determined by the corresponding effective refractive index.
For the decomposition, we simply consider the magnetic field of each eigenmode sampled in the center of the waveguides.
Momentum resolved spectra are obtained from the real space field distributions by a two-dimensional Fourier transform.
Figures \figref{Fig6}{a} and \figref{Fig6}{b} display the calculated real space intensity distribution and momentum-resolved spectrum, respectively, for a single site excitation in the bulk of the SSH model.
These calculations show the same features characterizing the experimental results, see Fig.~\figref{Fig2}{b} and \figref{Fig4}{b}.
In particular, we observe a ballistic spreading of the wave packet in the calculated real-space intensity distribution and two bands with slightly different spectral widths in the momentum resolved spectrum.
Analogously, the calculated real and Fourier space data for the case of the excitation at the boundary between the two dimerizations of the SSH model (excitation site I) are presented in Figs.~\figref{Fig6}{c} and \figref{Fig6}{d}, respectively.
As in the experiment, in addition to the bulk states we observe a localized mode at the boundary in the real space intensity distribution and the additional mode in the midgap position in the momentum-resolved spectra. 
We note that our calculation of SPP evolution does not assume nearest neighbor coupling, but takes into account all couplings terms in the array.
Our analysis shows a nonvanishing next-nearest neighbor coupling term of the same magnitude as that observed in the experiment, see Fig.~\figref{Fig4}{b,d}.
This gives independent confirmation of our assumption that the observed deviations from the ideal SSH model's prediction originates from a nonvanishing next-nearest neighbor coupling.
Its contribution can be further suppressed by increasing $a$.
However, this demands fabricating next generation SPP waveguide arrays with significantly reduced absorption losses, for example, using Ag instead of Au or using longer wavelengths \cite{Drezet:2008,Ditlbacher:2005ey,Ma:2010te}.

In conclusion, we have experimentally demonstrated the existence of a topologically protected edge state at the interface connecting the two topological phases of the SSH model.
To provide strong experimental evidence of its topological nature, we have shown through both real space and momentum-resolved measurements that the edge state has a midgap energy and populates only one sublattice, as predicted by theory.
Our work shows that combining real space with momentum resolved imaging gives new valuable physical insight that cannot be obtained from one domain only.
In future work, it would be interesting to study the topological protection against different classes of deformations of the system's parameters.

\begin{acknowledgments}
We are indebted to Wolfgang Alt, Janos A.~Asb\'oth, and Dieter Meschede for insightful discussions.
We acknowledge financial support from the Deutsche Forschungsgemeinschaft SFB/TR 185 OSCAR, and from the ERC Grant DQSIM (project ID: 291401).
\end{acknowledgments}

\bibliographystyle{apsrev4-1}
\bibliography{SSH-model-SPP}

\end{document}